\documentclass{emulateapj}
\usepackage{apjfonts}

\newcommand{\pivec}{\mbox{\boldmath $\pi$}}
\newcommand{\muvec}{\mbox{\boldmath $\mu$}}

\begin{document}

\title{Determining the Physical Lens Parameters of the Binary Gravitational 
Microlensing Event MOA-2009-BLG-016}

\author{
K.-H. Hwang\altaffilmark{1},
C. Han\altaffilmark{1,15,17},
I. A. Bond\altaffilmark{2,16},
N. Miyake\altaffilmark{3,16}\\
and\\
F. Abe\altaffilmark{3},
D.P. Bennett\altaffilmark{4},
C.S. Botzler\altaffilmark{5},
A. Fukui\altaffilmark{3},
K. Furusawa\altaffilmark{3},
F. Hayashi\altaffilmark{3},
J.B. Hearnshaw\altaffilmark{6},
S. Hosaka\altaffilmark{3},
Y. Itow\altaffilmark{3},
K. Kamiya\altaffilmark{3},
P.M. Kilmartin\altaffilmark{7},
A. Korpela\altaffilmark{8},
W. Lin\altaffilmark{2},
C.H. Ling\altaffilmark{2},
S. Makita\altaffilmark{3},
K. Masuda\altaffilmark{3},
Y. Matsubara\altaffilmark{3},
Y. Muraki\altaffilmark{9},
K. Nishimoto\altaffilmark{3},
K. Ohnishi\altaffilmark{10},
Y.C. Perrott\altaffilmark{5},
N. Rattenbury\altaffilmark{5},
To. Saito\altaffilmark{11},
T. Sako\altaffilmark{3},
L. Skuljan\altaffilmark{2},
D.J. Sullivan\altaffilmark{8},
T. Sumi\altaffilmark{3},
D. Suzuki\altaffilmark{3},
W.L. Sweatman\altaffilmark{2},
P.,J. Tristram\altaffilmark{7}, 
K. Wada\altaffilmark{9},
P.C.M. Yock\altaffilmark{5}\\
(The MOA Collaboration),\\
D.\ L., Depoy\altaffilmark{12},
%S.\ Dong\altaffilmark{13,14},
B.\ S.\ Gaudi\altaffilmark{13},
A.\ Gould\altaffilmark{13},
C.-U.\ Lee\altaffilmark{14},
%B.-G.\ Park\altaffilmark{14},
R.\ W.\ Pogge\altaffilmark{13}\\
(The $\mu$FUN Collaboration)\\
}

\altaffiltext{1}{Department of Physics, Chungbuk National University, Cheongju 361-763, Republic of Korea}
\altaffiltext{2}{Institute of Information and Mathematical Sciences, Massey University, Private Bag 102-904, North Shore Mail Centre, Auckland, New Zealand}
\altaffiltext{3}{Solar-Terrestrial Environment Laboratory, Nagoya University, Nagoya, 464-8601, Japan}
\altaffiltext{4}{Department of Physics, University of Notre Damey, Notre Dame, IN 46556, USA}
\altaffiltext{5}{Department of Physics, University of Auckland, Private Bag 92019, Auckland, New Zealand}
\altaffiltext{6}{University of Canterbury, Department of Physics and Astronomy, Private Bag 4800, Christchurch 8020, New Zealand}
\altaffiltext{7}{Mt. John Observatory, P.O. Box 56, Lake Tekapo 8770, New Zealand}
\altaffiltext{8}{School of Chemical and Physical Sciences, Victoria University, Wellington, New Zealand}
\altaffiltext{9}{Department of Physics, Konan University, Nishiokamoto 8-9-1, Kobe 658-8501, Japan}
\altaffiltext{10}{Nagano National College of Technology, Nagano 381-8550, Japan}
\altaffiltext{11}{Tokyo Metropolitan College of Industrial Technology, Tokyo 116-8523, Japan}
\altaffiltext{12}{Department of Physics, Texas A\&M University, College Station, TX, USA}
\altaffiltext{13}{Department of Astronomy, Ohio State University, 140 W. 18th Ave., Columbus, OH 43210, USA}
%\altaffiltext{14}{Institute for Advanced Study, School of Natural Sciences, Einstein Drive, Princeton, NJ 08540, USA}
\altaffiltext{14}{Korea Astronomy and Space Science Institute, Daejeon 305-348, Korea}
\altaffiltext{15}{Microlensing Follow Up Network ($\mu$FUN)}
\altaffiltext{16}{Microlensing Observations in Astrophysics (MOA) Collaboration}
\altaffiltext{17}{corresponding author}

\begin{abstract}
We report the result of the analysis of the light curve of 
the microlensing event MOA-2009-BLG-016. The light curve is 
characterized by a short-duration anomaly near the peak and 
an overall asymmetry.  We find that the peak anomaly is due 
to a binary companion to the primary lens and the asymmetry 
of the light curve is explained by the parallax effect caused 
by the acceleration of the observer over the course of the 
event due to the orbital motion of the Earth around the Sun. 
In addition, we detect evidence for the effect of the finite 
size of the source near the peak of the event, which allows 
us to measure the angular Einstein radius of the lens system.  
The Einstein radius combined with the microlens parallax allows 
us to determine the total mass of the lens and the distance to 
the lens.  We identify three distinct classes of degenerate 
solutions for the binary lens parameters, where two are 
manifestations of the previously identified degeneracies of 
close/wide binaries and positive/negative impact parameters, 
while the third class is caused by the symmetric cycloid shape 
of the caustic.  We find that, for the best-fit solution, the 
estimated mass of the lower-mass component of the binary is 
$(0.04 \pm 0.01)\ M_\odot$, implying a brown-dwarf companion. 
However, there exists a solution that is worse only by $\Delta 
\chi^2 \sim 3$ for which the mass of the secondary is above the 
hydrogen-burning limit. Unfortunately, resolving these two 
degenerate solutions will be difficult as the relative lens-source 
proper motions for both are similar and small 
($\sim 1~{\rm mas~yr^{-1}}$) and thus the lens will remain blended
with the source for the next several decades.
\end{abstract}

\keywords{gravitational lensing}

\section{Introduction}

For general microlensing events, the only lensing parameter that 
provides information about the physical parameters of the lens 
is the Einstein time scale $t_{\rm E}$.  However, the time scale 
results from the combination of the underlying physical lens  
parameters of the lens mass $M$, relative lens-source parallax 
$\pi_{\rm rel}={\rm AU}\ (D_{\rm L}^{-1}-D_{\rm S}^{-1})$, 
and proper motion $\mu$ by 
\begin{equation}
t_{\rm E}={\theta_{\rm E}\over \mu};\qquad
\theta_{\rm E}=(\kappa M \pi_{\rm rel})^{1/2},
\label{eq1}
\end{equation}
where $\kappa=4G/(c^2 {\rm AU})$, 
$\theta_{\rm E}$ represents the angular Einstein radius, and 
$D_{\rm L}$ and $D_{\rm S}$ are the distances to the lens and 
source star, respectively.  As a result, it is difficult to 
uniquely determine the physical lens parameters from the time 
scale alone.

% Figure 1 ----------------------------------------------------
\begin{figure*}[th]
\epsscale{0.8}
\plotone{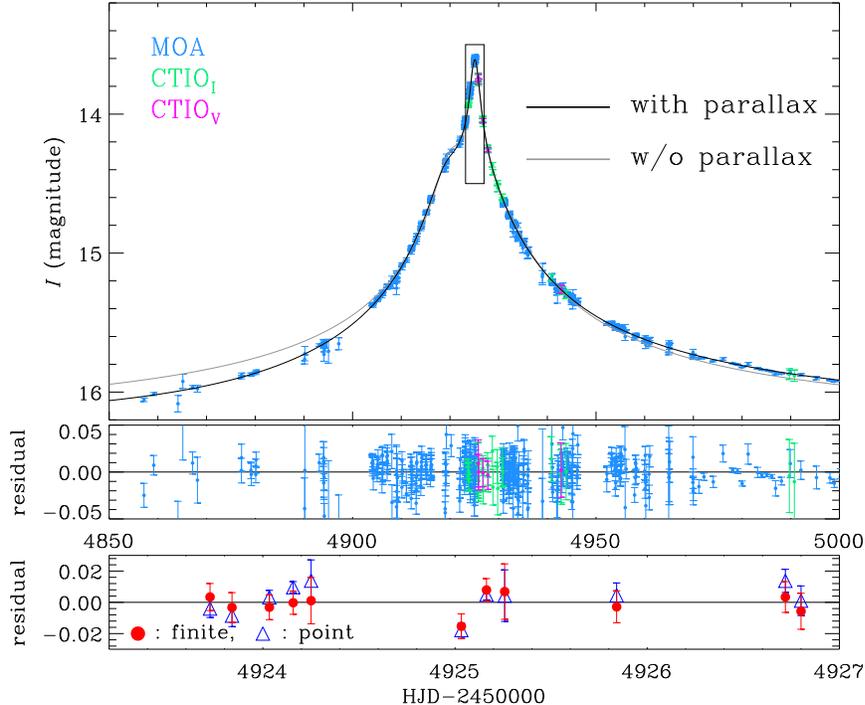}
\caption{\label{fig:one}
Light curve of the microlensing event MOA-2009-BLG-016. Also 
presented are the best-fit curves with and 
without the parallax effect.  Middle panel shows the residual 
from the best-fit model (both parallax and finite-source 
effect included).  The bottom panel shows the residuals from 
the best-fit finite-source (red) and point-source (blue) models 
in the region around the perturbation.  The portion of the light 
curve corresponding to the time span of the bottom panel is shown 
by a box in the upper panel.  We note that the data points in the 
bottom panel are binned by 4 hours to better show the difference. 
}\end{figure*}
% -------------------------------------------------------------

For complete determination of the physical parameters of a lens, 
it is required to measure both the lens parallax and angular 
Einstein radius.  The microlens parallax is defined by the ratio 
of the Earth's orbit to the Einstein radius projected on the 
observer plane, $\tilde{r}_{\rm E}$, i.e., 
\begin{equation}
\pi_{\rm E}={{\rm AU}\over \tilde{r}_{\rm E}}.
\label{eq2}
\end{equation}
In general, parallaxes are measured from the slight deviation of 
the overall shape of the light curve from a symmetric standard 
light curve \citep{paczynski86}, which is caused by the deviation 
of the relative lens-source motion from a rectilinear motion due 
to the orbital motion of the Earth around the Sun \citep{gould92}. 
Parallaxes are usually measured for long time-scale events for 
which the parallax effect is maximized.  Einstein radii, on the 
other hand, are generally measured from the deviation of the 
light curve caused by the finite size of source stars such as 
caustic-approaching events \citep{gould94}.  With the measured 
parallax and Einstein radius, the mass of the lens and the 
distance to the lens are uniquely determined by
\begin{equation}
M={\theta_{\rm E}\over \kappa \pi_{\rm E}},
\label{eq3}
\end{equation}
and
\begin{equation}
D_{\rm L}={{\rm AU}\over \pi_{\rm E}\theta_{\rm E}+\pi_{\rm S}},
\label{eq4}
\end{equation}
where $\pi_{\rm S}={\rm AU}/D_{\rm S}$ is the parallax of the 
source star.

Unfortunately, the conditions for the optimal measurements of 
the parallax and Einstein radius are different and thus the 
chance to completely determine the physical parameters of lenses 
by measuring both quantities is low.  In the literature, we find 
a total 13 microlensing events for which the lens masses were 
determined. These include 
EROS-2000-BLG-5 \citep{an02},
sc26-2218 \citep{smith03a},
OGLE-2002-BLG-018 \citep{kubas05},
OGLE-2003-BLG-235 \citep{bennett06},
OGLE-2003-BLG-238 \citep{jiang04},
OGLE-2006-BLG-109 \citep{gaudi08,bennett09},
OGLE-2007-BLG-050 \citep{batista09},
OGLE-2007-BLG-192 \citep{bennett08}
OGLE-2007-BLG-224 \citep{gould09},
OGLE-2008-BLG-279 \citep{yee09},
MACHO-LMC-5 \citep{alcock01,gould04},
OGLE-2003-BLG-175/MOA-2003-BLG-045 \citep{ghosh04}, and
OGLE-2005-BLG-071 \citep{udalski05, dong09}.

In this paper, we report the result of the analysis of the 
microlensing event MOA-2009-BLG-016.  We determine the mass 
of the lens and distance to it by measuring both the Einstein 
radius and lens parallax.  We identify three distinct types 
of degeneracy, where two are previously known and the other 
is newly identified in this work.

\section{Observation}

The microlensing event MOA-2009-BLG-016 occurred on a star located 
at $({\rm RA},{\rm DEC})=(17^\circ 57' 32.08'',-34^{\rm h}
21^{\rm m} 10.06^{\rm s})$, which corresponds to the Galactic 
coordinates of $(l,b)=(-4.1^\circ,-3.7^\circ)$.  It was first 
detected by the MOA collaboration on 2009 February 18 by using 
the 1.8 m telescope of Mt.\ John Observatory in New Zealand.  
An anomaly was detected on April 2 and it was announced to the 
microlensing community.  In response to the alert, the $\mu$FUN 
team conducted follow-up observations by using the 1.3 m SMARTS 
telescope of CTIO in Chile.  The CTIO data are composed of 29 
images in $I$ band and 5 images in $V$ band.  Data set from 
additional observatories were either single-epoch or near baseline 
and hence have not been included for analysis.  Photometric 
reductions of the data were carried out by using the codes 
developed by the individual groups.

In Figure \ref{fig:one}, we present the light curve of the event. 
We note that the magnitude of the light curve is not calibrated.
This is due to the lack of calibrated comparison stars in the 
field due to severe blending.  However, the lensed star can be 
constrained from modeling combined with the color information 
obtained from the position of the source star on the color-magnitude 
diagram relative to the position of the center of clump giants in 
the field for which the de-reddened magnitude and color are well 
known.  See more details in section 3.1.  We also note that data 
set from different observatories and filters are aligned by fitting 
them a common model.  \footnote{Photometric data used for analysis 
are available upon requests.}

\section{Characterization}

\subsection{Modeling}

Modeling microlensing light curves requires to include various 
parameters. To describe light curves of standard single-lens events, 
a set of three parameters are needed.  These include the Einstein 
time scale, $t_{\rm E}$, the time of the closest lens-source approach, 
$t_0$, and the lens-source separation normalized by the Einstein 
radius at the time of the closest approach, $u_0$.  If light curves 
exhibit binary-induced anomalies, an additional set of parameters 
is needed.  These parameters include the mass ratio between the 
lens components, $q$, the projected binary separation in units of 
the Einstein radius, $s$, and the angle of the source trajectory 
with respect to the binary axis, $\alpha$.  For many cases of binary
events, the normalized source radius, $\rho_\star\equiv \theta_\star/ 
\theta_{\rm E}$, is needed to describe the lensing magnification 
whenever the angular radius of the source star, $\theta_\star$, 
plays an important role such as in the vicinity of a caustic crossing.  
For some long time-scale events with parallax-induced deviations, 
it is required to include the parallax parameters $\pi_{{\rm E},N}$ 
and $\pi_{{\rm E},E}$, which are the components of the lens-parallax 
vector $\pivec_{\rm E}$ projected on the sky in the north and east 
celestial coordinates, respectively.  The direction of this vector 
is that of the lens-source relative motion in the frame of the Earth 
at the peak of the event.
Similar to the deviation by the parallax effect, the source trajectory 
can also be affected by the orbital motion of the source if it is 
composed of binary stars (`xallarap' effect).  Under the assumption 
of a circular orbit and a very faint binary companion, the xallarap 
effect is parameterized by the orbital period $P$, inclination $i$, 
and phase angle $\psi$ of the orbit.

Modeling light curves of microlensing events is a difficult task.
The large number of parameters makes brute-force searches of 
solutions difficult.  In addition, a simple downhill approach in 
the complicated $\chi^2$ surface often results in wrong solutions 
of local minima.  We, therefore, use a hybrid approach where grid 
searches are conducted over the space of the parameters of $s$, $q$, 
and $\alpha$ and the remaining parameters are searched by letting 
them vary so that they result in minimum $\chi^2$ at each grid of 
$s$, $q$, and $\alpha$ parameters.  We use a Markov Chain Monte 
Carlo method for $\chi^2$ minimization.  Once the $\chi^2$ minima 
of the individual grid points are determined, the best-fit model 
is obtained by comparing the $\chi^2$ values of the individual 
grids.  We investigate degeneracy of the solutions by probing 
local minima that appear in the space of the grid parameters.

% Figure 2 ----------------------------------------------------
\begin{figure}[ht]
\epsscale{1.1}
\plotone{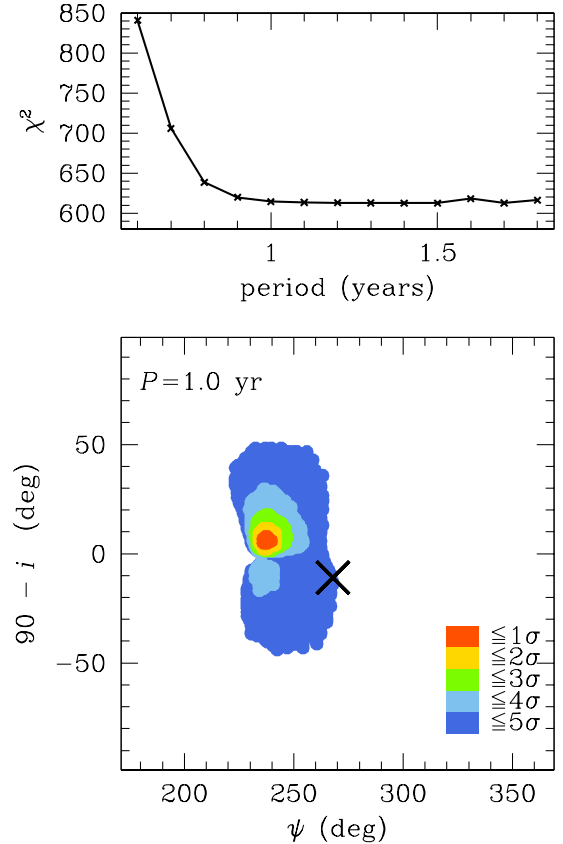}
\caption{\label{fig:two}
Distribution of $\Delta\chi^2$ with respect to the {\it xallarap}
parameters.  The upper panel shows the distribution as a function 
the binary-source orbital period, $P$, and the lower panel shows 
the distribution with respect to the orbital phase, $\psi$, and 
inclination, $i$ at a fixed orbital period of $P=1$ yr.  The position 
marked by ``X'' represents the position of the best-fit xallarap 
solution.
}\end{figure}
% -------------------------------------------------------------

The light curve of the event MOA-2009-BLG-016 is characterized by 
two important features.  One is the anomaly near the peak of the 
light curve and the other is the asymmetry of the overall light 
curve.  From modeling, we find that the anomaly near the peak is 
well explained by the central perturbation caused by a close/wide 
binary.  For the best-fit model, the determined values of the 
projected separation and mass ratio between the binary components 
are 
\begin{equation}
s=0.21\pm 0.01; \qquad q=0.33\pm 0.02.
\label{eq5}
\end{equation}
We also find that the asymmetry of the light curve can be explained 
by the parallax effect.  The determined values of the parallax 
parameters for the best-fit parallax model are 
\begin{equation}
\pi_{{\rm E},E}=0.230\pm 0.010;\qquad
\pi_{{\rm E},N}=0.108\pm 0.005.
\label{eq6}
\end{equation}
The improvement of the fit with the addition of the parallax effect 
is $\Delta\chi^2=3641$.  The parallax interpretation is consistent 
with the long time scale of the event, which is $t_{\rm E}\sim 135$ 
days for the best-fit model.

% Figure 3 ----------------------------------------------------
\begin{figure*}[ht]
\epsscale{0.8}
\plotone{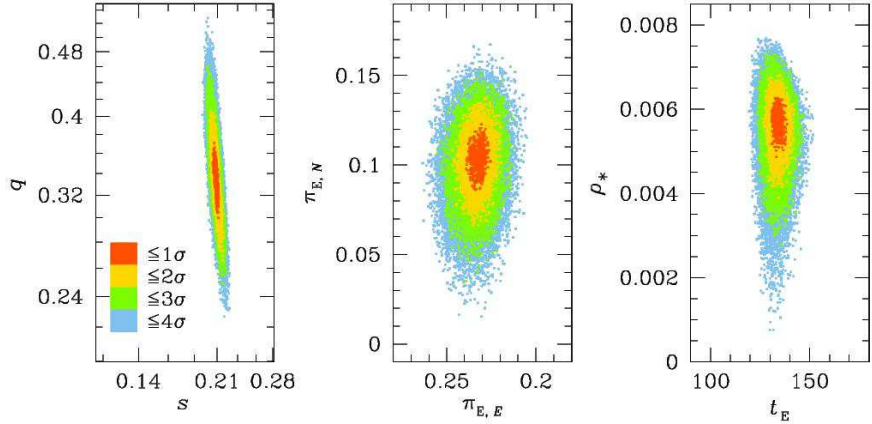}
\caption{\label{fig:three}
Contours of $\Delta\chi^2$ in the spaces of the binary parameters 
$(s,q)$ [left panel], the parallax parameters $(\pi_{{\rm E},N}, 
\pi_{{\rm E},E})$ [middle panel], and the normalized source radius 
and the Einstein time scale $(\rho_\star,t_{\rm E})$ [right panel] 
in the neighborhood of the best-fit solution (close I-- model).
}\end{figure*}
% -------------------------------------------------------------

It is known that a parallax signal can be mimicked by that of 
the xallarap effect \citep{smith03b}.   To check this possibility,  
we search for xallarap solutions under the assumption that the 
binary source is in a circular orbit.  Figure \ref{fig:two} shows 
the distributions of $\Delta\chi^2$ as a function of the xallarap 
parameters obtained from modeling.

We find that the best fit xallarap solution yields an improvement 
of $\Delta\chi^2=29$ for 3 additional degrees of freedom.  Although 
this improvement is formally highly significant and certainly implies 
that the 3 additional xallarap parameters are capable of ``responding'' 
to systematic deviations in the data that are not fully captured by 
the parallax modeling, we will now argue that the parallax solutions 
is preferred and the additional systematic deviations ``detected'' 
by the xallarap parameters are most likely due to other effects.

The first point is that the parallax signal ($\Delta\chi^2=3641$)
is more than 100 times stronger than the additional signal from
xallarap, and such 1\% systematic effects are quite common in
microlensing events.  These may be due to real effects, such
as binary orbital motion, third bodies in the system, or other
physical effects, or may simply be due to systematics in the
data, which (because of the fleeting nature of the events) are
taken under an extremely wide range of conditions.  Systematic
effects at this level can never be definitively tracked down
because the $\Delta\chi^2$ is too small to adequately characterize
the real effects (such as the mass and separation of a putative
third body).

Second, if we consider the entire two-dimensional space of
possible binary orientations, only 1/6 lie within $\Delta\chi^2<29$
of the minimum, i.e., as close as the parallax solution (see Figure 
\ref{fig:two}).  If we further account for the fact that all events 
must have parallax at some level, whereas only half of sources have 
a binary companion of any mass, with period $P>1\,{\rm yr}$ (see 
\ref{fig:two}), then the prior probability of a xallarap solution 
is only 1/12.  If there were known to be no other systematic effects 
(real or instrumental) affecting this event, then this 1/12 probability
would have meager weight against a $\Delta\chi^2=29$.  But since such 
low-level systematics are in fact common,  the 1/12 probability must 
be taken seriously.  In brief, while we cannot rule out the xallarap 
solution, we judge it to be relatively unlikely and so adopt the 
parallax solution.

We find that a finite-source model is preferred over a point-source 
model with $\Delta\chi^2=12$.  The amount of $\Delta\chi^2$ is not 
big, but we note that it is statistically significant considering 
that the signal of the finite-source effect lasts only a short period 
of time during the source's approach close to the cusp of the caustic.

In Figure~\ref{fig:one}, we 
present the best-fit model curve on the top of the light curve.  
Also presented in the top panel is the model curve without the 
parallax effect.  Middle panel shows the residual from the best-fit 
model (both the parallax and finite-source effect included).  The 
bottom panel shows the residuals from the best-fit finite-source 
(red) and point-source (blue) models in the region around the 
perturbation.  We note that the data points in the bottom panel 
are binned by 4 hours to better show the difference.

In Figure~\ref{fig:three}, we present the contours of $\Delta\chi^2$ 
in the spaces of the binary parameters $(s,q)$ [left panel], the 
parallax parameters $(\pi_{{\rm E},E}, \pi_{{\rm E},N})$ [middle 
panel], and the normalized source radius and the Einstein time scale 
$(\rho_\star, t_{\rm E})$ [right panel] in the neighborhood of the 
best-fit model.

% Figure 4 ----------------------------------------------------
\begin{figure}[t]
\epsscale{1.1}
\plotone{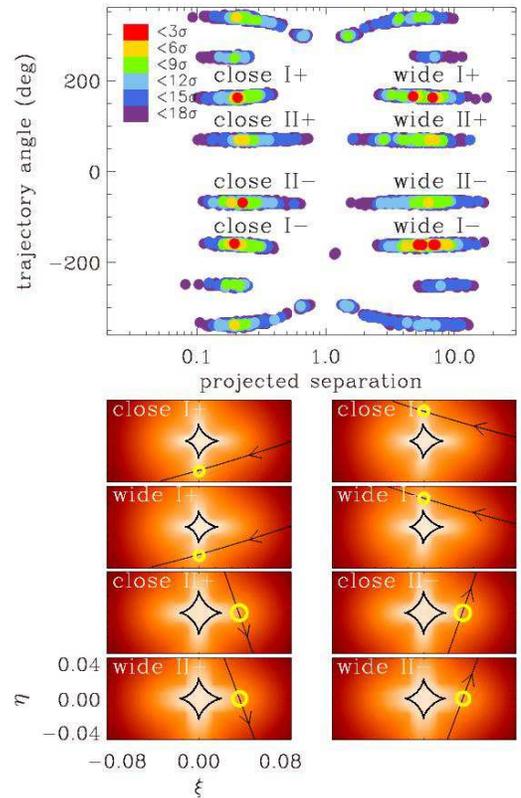}
\caption{\label{fig:four}
Likelihood contours for binary solutions as a function of the 
projected binary separation (normalized by the Einstein radius) 
and the source trajectory angle (upper panel).  Panels in the 
lower half show the geometry of the source trajectory (straight 
line with an arrow) with respect to the position of the caustic 
for the solutions corresponding to the individual local minima.  
The small yellow circle on the source trajectory represents the 
source star at the moment of the closest caustic approach.  
Among the 8 pairs of the local minima, we label only 4 pairs 
for which the values of $\Delta\chi^2$ from that of the best-fit 
model are relatively small.
}\end{figure}
% -------------------------------------------------------------

\begin{deluxetable*}{lrrrrrrrr}[h]
\tablecaption{Fit Parameters\label{table:one}}
\tablewidth{0pt}
\tablehead{
\multicolumn{1}{c}{parameters} &
\multicolumn{1}{c}{close I+} & 
\multicolumn{1}{c}{{\bf close I--}} & 
\multicolumn{1}{c}{wide I+} & 
\multicolumn{1}{c}{wide I--} & 
\multicolumn{1}{c}{close II+} & 
\multicolumn{1}{c}{close II--} & 
\multicolumn{1}{c}{wide II+} & 
\multicolumn{1}{c}{wide II--} 
}
\startdata
% ----------------------------------------------------------------------------------------------------
$\chi^2$/dof     & 649.1/645 & {\bf 642.8/645 } & 655.9/645      &  645.7/645      &  663.7/645 &  669.6/645 &  673.0/645      &  673.8/645      \\ 
$s$              & 0.211     & {\bf 0.208 }     & 7.190          &  7.296          &  0.210     &  0.207     &  9.099          &  8.934          \\
$q$              & 0.298     & {\bf 0.333 }     & 0.686          &  0.735          &  0.582     &  0.610     &  5.157          &  4.002          \\
$\alpha$ (deg)   & 163.20    & {\bf -162.56 }   & 163.53         & -162.91         &  70.12     & -70.11     &  69.22          & -69.18          \\
$t_0$ (HJD')     & 4923.805  & {\bf 4923.779 }  & 4923.914       &  4923.882       &  4923.928  &  4923.934  &  4923.780       &  4923.785       \\
$u_0$            & 0.027     & {\bf -0.028 }    & 0.021(0.027)   & -0.020(-0.027)  &  0.036     & -0.036     &  0.015(0.036)   & -0.016(-0.036)  \\
$t_E$            & 135.77    & {\bf 135.29 }    & 173.21(133.40) &  177.17(134.50) &  106.24    &  108.55    &  274.62(110.68) &  247.69(110.75) \\
$\rho_\star$     & 0.0055    & {\bf 0.0056 }    & 0.0046(0.0059) &  0.0043(0.0057) &  0.0087    &  0.0085    &  0.0031(0.0076) &  0.0035(0.0079) \\
$\pi_{E,N}$      & 0.090     & {\bf 0.108 }     & 0.065(0.084)   &  0.082(0.108)   & -0.026     & -0.023     & -0.011(-0.028)  & -0.010(-0.021)  \\
$\pi_{E,E}$      & 0.224     & {\bf 0.230 }     & 0.171(0.221)   &  0.180(0.237)   &  0.275     &  0.268     &  0.104(0.257)   &  0.117(0.261)   \\
$f_{{\rm b},I} $ & 0.692     & {\bf 0.688 }     & 0.690          &  0.692          &  0.594     &  0.604     &  0.600          &  0.602          \\
$f_{{\rm b},V} $ & 0.693     & {\bf 0.694 }     & 0.690          &  0.697          &  0.604     &  0.613     &  0.604          &  0.608          
% ----------------------------------------------------------------------------------------------------
\enddata
\tablecomments{
${\rm HJD}'={\rm HJD}-2450000$. The parameters of the best-fit solution 
are marked in bold fonts.  The parameters in the parentheses for wide 
solutions represent the values with respect to the mass of the binary
component associated with the caustic involved with the perturbation. 
We note that $f_{{\rm b},V}$ and $f_{{\rm b},I}$ represent the blended 
light fractions in $V$ and $I$ passbands, respectively, measured from 
the CTIO data.
}
\end{deluxetable*}

\subsection{Degeneracy}

Although the basic characteristics of the lens system are defined, 
we find that there exist degenerate solutions.  The upper panel 
of Figure~\ref{fig:four} shows the local minima in the parameter 
space of the projected binary separation and the source trajectory 
angle.  Panels in the lower half show the geometry of the source 
trajectory with respect to the caustic for the solutions corresponding 
to the individual local minima.  We note that among the total 8 
pairs of local minima we label only 4 pairs for which the values 
of $\Delta\chi^2$ from the best-fit model are relatively small.
From the analysis of the individual local minima, we find that 
they result from three distinct types of degeneracy.

The first type of degeneracy is the well-known close/wide 
degeneracy \citep{dominik99,albrow99,afonso00,albrow02}.  This 
degeneracy is caused by the similarity in shape between the 
caustics induced by a wide-separation binary with $s>1$ and a 
close-separation binary with $s<1$.  This can be seen in 
Figure~\ref{fig:four}, where one finds pairs of local minima 
with separations $s>1$ and $s<1$.

The second type of degeneracy is caused by the mirror symmetry 
between the pair of source trajectories with the impact parameters
and source trajectory angles of $(u_0,\alpha)$ and $(-u_0, -\alpha)$ 
\citep{smith03b}.  For a rectilinear motion, the two light curves 
resulting from the two source trajectories are identical.  If the 
parallax effect is not negligible, however, the light curves from 
the trajectories are slightly different due to the curvature of the 
source trajectory.  The pairs of local minima with trajectory angles 
$\alpha$ and $-\alpha$ in Figure~\ref{fig:four} are caused by this 
degeneracy.

The third type of degeneracy is caused by the shape of the caustic. 
When the caustic is produced by a binary with a separation significantly 
larger or smaller than the Einstein radius, its shape is a symmetric 
cycloid with four cusps where two cusps are on the binary axis and 
the others are off the axis.  Then, source trajectories approaching 
the caustic with angles $\alpha$, $\alpha+\pi/2$, $\alpha+\pi$, and 
$\alpha+3\pi/2$ result in a similar perturbation.  We refer to this 
degeneracy as `cycloid degeneracy'.  We find that the caustic responsible 
for the central perturbation of the event MOA-2009-BLG-016 is very 
symmetric and thus the light curve is subject to this degeneracy.  
The four local minima on each quadrant of $(s,\alpha)$ parameter space 
in Figure~\ref{fig:three} are caused by this degeneracy.

In Table~\ref{table:one}, we list the lensing parameters of the 
local minima along with values of $\chi^2$.  We find that the 
models with trajectory angle of $|\alpha|\sim 163^{\circ}$ provide 
better fits than the corresponding models with $|\alpha|\sim 70^{\circ}$.  
Among the close-wide pairs of solutions with $|\alpha|\sim 163^{\circ}$, 
we find close-binary models are preferred.  Among the two close-binary 
models with $|\alpha|\sim 163^{\circ}$, we find that the ``$-u_0$'' 
model (close I--) provides the best fit to the observed light curve.

\begin{deluxetable*}{lrrrcccccc}[h]
\tablecaption{Physical Parameters\label{table:two}}
\tablewidth{0pt}
\tablehead{
\multicolumn{1}{c}{model} &
\multicolumn{1}{c}{$\Delta\chi^2$} &
\multicolumn{1}{c}{$\theta_{\rm E}$} & 
\multicolumn{1}{c}{$\mu$} & 
\multicolumn{1}{c}{$\mu_\odot$} & 
\multicolumn{1}{c}{$\varphi_\mu$} & 
\multicolumn{1}{c}{$D_{\rm L}$} & 
\multicolumn{1}{c}{$M$} &
\multicolumn{1}{c}{$M_1$} &
\multicolumn{1}{c}{$M_2$} \\
\multicolumn{1}{c}{} &
\multicolumn{1}{c}{} &
\multicolumn{1}{c}{(mas)} & 
\multicolumn{1}{c}{(mas yr$^{-1}$)} & 
\multicolumn{1}{c}{(mas yr$^{-1}$)} & 
\multicolumn{1}{c}{(deg)} & 
\multicolumn{1}{c}{(kpc)} & 
\multicolumn{1}{c}{($M_{\odot}$)} &
\multicolumn{1}{c}{($M_{\odot}$)} &
\multicolumn{1}{c}{($M_{\odot}$)} 
}
\startdata
% ------------------------------------------------------------------------
close I+        & 6.3       & 0.36$\pm$0.07                 & 0.97$\pm$0.17       & 1.12$\pm$0.20       & 65.9       & 4.80$\pm$0.60       & 0.18$\pm$0.03       & 0.14$\pm$0.03       & 0.04$\pm$0.01       \\
{\bf close I--} & {\bf 0.0} & {\bf 0.36$\pm$0.06}           & {\bf 0.96$\pm$0.17} & {\bf 1.13$\pm$0.20} & {\bf 63.1} & {\bf 4.70$\pm$0.58} & {\bf 0.17$\pm$0.03} & {\bf 0.13$\pm$0.02} & {\bf 0.04$\pm$0.01} \\
wide I+         & 13.1      & 0.43$\pm$0.08 (0.33$\pm$0.06) & 0.91$\pm$0.17       & 1.05$\pm$0.20       & 66.9       & 4.98$\pm$0.64       & 0.29$\pm$0.06       & 0.17$\pm$0.03       & 0.12$\pm$0.02       \\
wide I--        & 2.9       & 0.46$\pm$0.08 (0.35$\pm$0.06) & 0.94$\pm$0.16       & 1.11$\pm$0.19       & 63.6       & 4.72$\pm$0.59       & 0.28$\pm$0.05       & 0.16$\pm$0.03       & 0.12$\pm$0.02       \\
close II+       & 21.0      & 0.27$\pm$0.03                 & 0.91$\pm$0.10       & 1.02$\pm$0.11       & 90.1       & 5.12$\pm$0.58       & 0.12$\pm$0.01       & 0.08$\pm$0.01       & 0.04$\pm$0.01       \\
close II--      & 26.8      & 0.27$\pm$0.03                 & 0.90$\pm$0.11       & 1.00$\pm$0.12       & 89.7       & 5.16$\pm$0.60       & 0.12$\pm$0.02       & 0.08$\pm$0.01       & 0.05$\pm$0.01       \\
wide II+        & 30.3      & 0.74$\pm$0.14 (0.30$\pm$0.06) & 0.99$\pm$0.19       & 1.10$\pm$0.21       & 90.9       & 5.01$\pm$0.67       & 0.88$\pm$0.19       & 0.14$\pm$0.03       & 0.73$\pm$0.16       \\
wide II--       & 31.0      & 0.64$\pm$0.11 (0.29$\pm$0.05) & 0.95$\pm$0.16       & 1.06$\pm$0.18       & 89.6       & 5.07$\pm$0.65       & 0.68$\pm$0.13       & 0.14$\pm$0.03       & 0.54$\pm$0.11       
% -----------------------------------------------------------------------
\enddata
\tablecomments{
Here $\theta_{\rm E}$ represents the angular Einstein radius,
$\mu=\theta_{\rm E}/t_{\rm E}$ and $\mu_\odot$ represent the 
geocentric and heliocentric lens-source proper motion, respectively, 
$\varphi_\mu$ is the angle of the proper motion with respect to 
the east, $D_{\rm L}$ is the distance to the lens, $M$ is the total 
mass of the binary lens, and $M_1$ and $M_2$ are the masses of the 
binary components.  We note that the subscript ``1'' is used to 
denote the lens component located closer to the source trajectory 
and thus $M_1$ can be smaller than $M_2$.  The parameters of the 
best-fit solution are marked in bold fonts.  The values of 
$\Delta\chi^2$ are with respect to the best-fit solution, i.e.\ 
close I-- model.  The Einstein radius in parenthesis represents 
the value with respect to the mass of the binary component 
associated with the caustic involved with the perturbation.
}
\end{deluxetable*}

% Figure 5 ----------------------------------------------------
\begin{figure}[h]
\epsscale{1.1}
\plotone{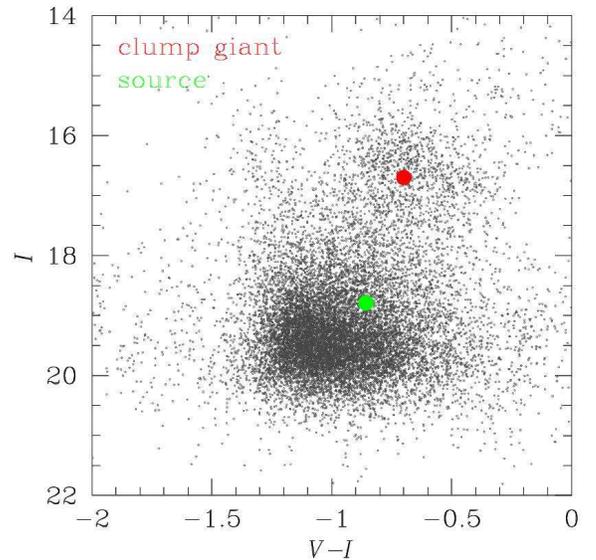}
\caption{\label{fig:five}
Position of the source (lensed) star with respect to the centroid of clump
giants in the instrumental (uncalibrated) color-magnitude diagram.
}\end{figure}
% -------------------------------------------------------------

\subsection{Physical Parameters}

To determine $\theta_\star$, we first determine the de-reddened 
magnitude $I_0$ and color $(V-I)_0$ of the source star from the 
measured offset between the source and the centroid of the clump 
giants in the instrumental color-magnitude diagram (CMD) constructed 
by using the $V$ and $I$ band images taken from CTIO 
(Figure~\ref{fig:five}) under the assumption that the source star 
and clump giants experience the same amount of extinction.  The 
locations of the source star and the centroid of clump giants on 
the instrumental CMD are $(V-I,I)_{\rm S}= (-0.86, 18.79)$ and 
$(V-I,I)_{\rm C}= (-0.70, 16.70)$, respectively.  The location of 
the source on the CMD is determined based on the light fractions 
of the source star in the $I$ and $V$ bands determined from modeling.
With the known de-reddened magnitude and color of bulge clump 
giants of $[(V-I)_0,I_0]_{\rm C}=(1.04,14.32)$ toward the field, 
the de-reddened brightness and color of the source star are
determined by
$(V-I)_{0,{\rm S}}=[(V-I)_{\rm S}-(V-I)_{\rm C}]+1.05=0.89$ 
and $I_{0,{\rm S}}=(I_{\rm S}-I_{\rm C})+14.52 = 16.41$,
respectively.  Here we adopt a mean distance to clump giants 
toward the field of 8.8 kpc estimated by \citet{rattenbury07}.
Then, the angular source size is determined by first transforming 
from $(V-I)_0$ to $(V-K)_0$ using the color-color relation of 
\citet{bessel98} and then applying the relation between $(V-K)_0$ 
and the angular stellar radius of \citet{kervella04}.  The resulting 
angular radius is 
\begin{equation}
\theta_\star=(2.00\pm 0.20)\ \mu{\rm as},
\label{eq7}
\end{equation}
where the uncertainty is estimated from the combination of the 
uncertainties of the colors and magnitudes of the source star 
and and an additional 7\% intrinsic error in the conversion 
process from the measured color to source radius \citep{yee09}.  
With the measured source radius, the Einstein radius and lens-source 
proper motion are determined by 
\begin{equation}
\theta_{\rm E} ({\rm close\ I-}) = \theta_\star /\rho_\star
= (0.36\pm0.06)\ {\rm mas}
\label{eq8}
\end{equation}
and
\begin{equation}
\mu ({\rm close\ I-}) = \theta_{\rm E}/t_{\rm E}
= (0.96 \pm 0.17)\ {\rm mas}\ {\rm yr}^{-1},
\label{eq9}
\end{equation}
respectively.

With the measured parallax and Einstein radius, the mass of 
the lens system and distance to the lens are determined from 
the relations in equation (3) and (4).  For the best-fit model, 
these values are
\begin{equation}
M  ({\rm close\ I-}) = (0.17 \pm 0.03)\ M_\odot
\label{eq10}
\end{equation}
and 
\begin{equation}
D_{\rm L} ({\rm close\ I-}) = (4.70\pm 0.58)\ {\rm kpc},
\label{eq11}
\end{equation}
respectively.  For the estimation of the uncertainty of $D_{\rm L}$,
we consider an $17\%$ fractional error of the source location 
estimated by the bulge mass distribution model of \citet{han95}.  
The values for other solutions are presented in Table \ref{table:two}.  
We note that, for the best solution, the estimated mass of the 
lower-mass component of the binary is 
\begin{equation}
M_2\ ({\rm close\ I-})=0.04\pm 0.01\ M_\odot,
\label{eq12}
\end{equation}
making it a brown-dwarf candidate.  We note, however, that the 
estimated mass of the companion for the second-best model 
is above the hydrogen-buring limit.

\subsection{Resolution of Degeneracies}

It is found that analysis of the light curve of MOA-2009-BLG-016 
alone results in degenerate solutions.  However, it might be possible 
to resolve the degeneracies with extra information.  We check the 
possibility of resolving the degeneracies from the measurement of 
proper motion from high-resolution observations.  For this, we 
compute the heliocentric lens-source proper motion $\muvec_\odot$ 
for the individual solutions.

In Table \ref{table:one}, we present the magnitudes and directions 
of the proper motion vectors for the individual solutions.  From 
the table, it is found that the high-resolution observation would 
be of limited use due to several reasons.  First, the two sets of 
solutions caused by the cycloid degeneracy result in different 
directions of proper motion and thus resolution of the lens and 
source would make it possible to resolve the degeneracy, but this 
degeneracy is already clearly lifted with significant $\Delta\chi^2$ 
from the light curve alone.  Moreover, this would be only possible 
many years after the event considering the small magnitude of the 
proper motion of $\mu_\odot \sim  1\ {\rm mas}\ {\rm yr}^{-1}$.  
Second, the solutions caused by other degeneracies result in proper 
motions with not only a similar magnitude but also a similar direction, 
implying that the degeneracies would be difficult to be lifted even 
with high-resolution 
observations.

\section{Conclusion}

We analyzed the light curve of a microlensing event MOA-2009-BLG-016,
which is characterized by high-magnification with an anomaly near 
the peak and an overall asymmetric light curve.  We found that the 
anomaly and asymmetry of the light curve are explained by the 
lens binarity and the parallax effect, respectively.  With the Einstein 
radius measured from the central perturbation combined with the lens 
parallax measured from the overall asymmetric light curve, we determined 
the mass of the lens and distance to the lens.  We identified three 
distinct types of degeneracy, where two were previously known and 
the other is first identified in this work.  We also found that, for 
the best solution, the estimated mass of the lower-mass component of 
the binary is in the mass range of brown dwarfs.

We acknowledge the following support: Creative Research Initiative 
Program (2009-0081561) of National Research Foundation of Korea (CH);
Korea Astronomy and Space Science Institute (B-GP, C-UL);
NSF AST-0757888 (AG); NASA  NNG04GL51G (BSG, AG, RWP); 
NSF AST-0708890 (DPB); NASA NNX07AL71G (BPB);
JSPS20740104 (TS).

\end{document}